\documentstyle[]{article}
\makeatletter
\newdimen\normalarrayskip              
\newdimen\minarrayskip                 
\normalarrayskip\baselineskip
\minarrayskip\jot
\newif\ifold             \oldtrue            
\def\arraymode{\ifold\relax\else\displaystyle\fi} 
\def\eqnumphantom{\phantom{(\theequation)}}     
\def\@arrayskip{\ifold\baselineskip\z@\lineskip\z@
     \else
     \baselineskip\minarrayskip\lineskip2\minarrayskip\fi}
\def\@arrayclassz{\ifcase \@lastchclass \@acolampacol \or
\@ampacol \or \or \or \@addamp \or
   \@acolampacol \or \@firstampfalse \@acol \fi
\edef\@preamble{\@preamble
  \ifcase \@chnum
     \hfil$\relax\arraymode\@sharp$\hfil
     \or $\relax\arraymode\@sharp$\hfil
     \or \hfil$\relax\arraymode\@sharp$\fi}}
\def\@array[#1]#2{\setbox\@arstrutbox=\hbox{\vrule
     height\arraystretch \ht\strutbox
     depth\arraystretch \dp\strutbox
     width\z@}\@mkpream{#2}\edef\@preamble{\halign \noexpand\@halignto
\bgroup \tabskip\z@ \@arstrut \@preamble \tabskip\z@ \cr}%
\let\@startpbox\@@startpbox \let\@endpbox\@@endpbox
  \if #1t\vtop \else \if#1b\vbox \else \vcenter \fi\fi
  \bgroup \let\par\relax
  \let\@sharp##\let\protect\relax
  \@arrayskip\@preamble}
\def\eqnarray{\stepcounter{equation}%
              \let\@currentlabel=\theequation
              \global\@eqnswtrue
              \global\@eqcnt\z@
              \tabskip\@centering
              \let\\=\@eqncr
              $$%
 \halign to \displaywidth\bgroup
    \eqnumphantom\@eqnsel\hskip\@centering
    $\displaystyle \tabskip\z@ {##}$%
    &\global\@eqcnt\@ne \hskip 2\arraycolsep
         $\displaystyle\arraymode{##}$\hfil
    &\global\@eqcnt\tw@ \hskip 2\arraycolsep
         $\displaystyle\tabskip\z@{##}$\hfil
         \tabskip\@centering
    &{##}\tabskip\z@\cr}
\makeatother

\def\beq{\begin{equation}}
\def\eeq{\end{equation}}
\def\bea{\begin{eqnarray}}
\def\eea{\end{eqnarray}}

\def\nn{\nonumber}

\begin{document}

\begin{titlepage}
\begin{center}
{{\it P.N.Lebedev Institute preprint} \hfill FIAN/TD-7/92\\
{\it I.E.Tamm Theory Department} \hfill ITEP-M-5/92\\
\hfill hepth@xxx/92\#\#
\begin{flushright}{July 1992}\end{flushright}
\vspace{0.1in}{\Large\bf Landau-Ginzburg Topological Theories\\
in the Framework of GKM and Equivalent Hierarchies}\\[.4in]
{\large  S. Kharchev, A. Marshakov, A. Mironov}\\
\bigskip {\it  P.N.Lebedev Physical
Institute \\ Leninsky prospect, 53, Moscow, 117 924, Russia},
\footnote{E-mail address: tdparticle@glas.apc.org   \&
mironov@sci.fian.msk.su}\\ \smallskip
\bigskip {\large A. Morozov}\\
 \bigskip {\it Institute of Theoretical and Experimental
Physics,  \\
 Bol.Cheremushkinskaya st., 25, Moscow, 117 259, Russia},
 \footnote{E-mail address: morozov@itep.msk.su}}
\end{center}
\bigskip \bigskip

\bigskip

\begin{abstract}
We consider the deformations of ``monomial solutions" to Generalized Kontsevich
Model \cite{KMMMZ91a,KMMMZ91b}
and establish the relation between the flows generated by
these deformations with those of
$N=2$ Landau-Ginzburg topological theories. We prove that the partition
function of a generic Generalized Kontsevich Model can be presented as a
product
of some ``quasiclassical" factor and non-deformed partition function which
depends only on the sum of Miwa transformed and flat times.
This result is important for the
restoration of explicit $p-q$ symmetry in the interpolation pattern between all
the $(p,q)$-minimal string models with $c<1$ and for revealing its integrable
structure in $p$-direction, determined by deformations of the potential.
It also implies the way in which supersymmetric
Landau-Ginzburg models are embedded into the general context of GKM. From the
point of view of integrable theory these deformations present a particular case
of what is called equivalent hierarchies.
\end{abstract}

\end{titlepage}

\newpage
\setcounter{footnote}0

\bigskip

{\bf 1.}  In \cite{KMMMZ91a,KMMMZ91b} a new model was introduced, which
naturally incorporates all
the non-perturbative partition functions of $c<1$ minimal string models and
parameterizes them by unique potential which allows one to interpolate smoothly
between the models.

On one hand, this potential if polynomial describes rolling among
different reductions of KP hierarchy so that the whole space of these
reductions
falls out into
different non-connected ``orbits" (or ``domains" or ``universality" classes)
marked by the higher
degree of the polynomial. Thus, one problem is of explicit description of
this rolling.

On the other hand, it seems that GKM has a universal nature, containing
information about {\it all} subjects related to 2-dimensional gravity. One
example of this phenomena has been already discussed in \cite{KMMM92a},
where it was
shown that discrete matrix models have a nice description in terms of GKM.
There is another interesting problem -- the relation between GKM and
topological Landau-Ginzburg Models (LGM)
\footnote{ In this letter by topological LGM we mean the LGM interacting
with topological gravity.} ,
which were the subject of a very
interesting recent development \cite{LVW89,Vaf90,DVV91b,Los92,Dij91}.

It turns out that the resolution to both above mentioned problems can be found
along the same line of investigation. Indeed, the connection to LGM is
naturally related to the question of interpolation between various
$(p,q)$-solutions to  $c<1\ \ 2d$ gravity \cite{KMMMZ91a,KMMMZ91b,Mar92},
$i.e.$ different
reductions of KP system, and quasiclassical limit of GKM. In this letter
we concentrate on the most important point in this connection  --  the
appearance and relation between corresponding integrable hierarchies which in
the language of integrable theories is nothing but what is called ``equivalent
hierarchies" \cite{Shi86}.

In order to do this, we first compute the derivatives with respect to first
several ``Miwa times" and show that they are naturally expressed through
corresponding derivatives with respect to the coefficients of the potential.
Moreover, it turns out that there are special combinations of these
coefficients --- so called flat or $p$-times \cite{DVV91b,Kri92,Dub91}
--- which naturally
occur in the framework of GKM
\footnote{We
call the particular linear combinations of the coefficients of the
potential $p$-times because they describe the deformation in one of possible
directions  ---
  ``$p$-direction" in deforming $(p,q)$-models, while deformation
in the other ``$q$-direction" is described by Miwa times. To avoid
misunderstanding, in \cite{Mar92} the opposite notations were used.
}
and are just those variables in what two above mentioned problems are
simultaneously resolved.

In the most elegant way all the formulas may be rewritten using
reparameterizations
of the spectral parameter, which naturally lead to the description in terms of
equivalent hierarchies. Different equivalent hierarchies in the GKM context
are parameterized by
polynomials of the same degree, and these polynomials are just superpotentials
of
LGM. Thus, we derive our main result --- that ``the topologically-deformed" GKM
is expressed through the non-deformed one, being equivalent solution to the
same integrable hierarchy (or, put differently, corresponding to different
reductions of KP hierarchy which lie at the same ``orbit").

To be more precise, first we remind that the partition function of GKM is
defined as a matrix integral \cite{KMMMZ91a,KMMMZ91b}
\beq\label{1}
Z^{(N)}_{GKM}[V|M] \equiv  C^{(N)}_{GKM}[V|M] e^{TrV(M)-TrMV'(M)}\int
DX\ e^{-TrV(X)+TrV'(M)X}
\eeq
over $N\times N$ ``Hermitean" matrices, with normalization factor given by
the Gaussian integral
\bea\label{2}
C^{(N)}_{GKM}[V|M]^{-1} \equiv  \int  DY \ e^{-TrV_2[M,Y]},\nn\\
V_2 \equiv  \lim_{\epsilon \rightarrow 0}
{1\over \epsilon ^2}Tr[V(M+\epsilon Y) - V(M) - \epsilon YV'(M)] .
\eea
The partition function $Z_{GKM}$ actually depends on  $M$  through the
invariant
variables
\beq\label{3}
T_k = {1\over k} Tr\ M^{-k}\hbox{, }  k\geq 1,
\eeq
moreover, if rewritten in terms of $T_k$,  $Z_{GKM}[V|T] = Z^{(N)}_{GKM}[V|M]$
is essentially independent of the size $N$ of the matrices: the only origin of
$N$-dependence is that of $T_k$ themselves (for finite $N$ not all the  $T_k$
are algebraically independent).

As a function of  $T_k$  $Z_{GKM}[V|T_k]$ is a  $\tau $-function of
KP-hierarchy,
$Z_{GKM}[V|T_k] = \tau _V[T_k]$, while ``potential"  $V$  specifies the
relevant point of the infinite-dimensional Grassmannian.

For various choices of the ``potential"  $V(X)$  $GKM$ reproduces continuum
limits of all multimatrix models:  $\displaystyle{V(X) = {X^{p+1}\over p+1}}$
is associated
with  the $(p-1)$-matrix model and thus with the entire set of $(p,q)$-minimal
string models with all possible $q$'s. In order to specify $q$ one needs to
make a special choice of $T$-variables: all $T_k= 0$, except for  $T_1$ and
$T_{p+q}$. The symmetry between $p$ and $q$ is implicit in this kind of
formulation and is to be revealed somehow in the future studies. However, it
immediately implies that since integrability occurs in $q$-direction ($i.e.$
$Z$  is a $\tau $-function as a function of $T$'s), it should also appear in
$p$-direction ($i.e.$ in some sense $Z$ should also be a $\tau $-function as a
function of the coefficients of the potential $V$).

In this letter we demonstrate that this is indeed the case, moreover, that the
$\tau $-function is essentially the same in both directions. Indeed, from the
considerations of the spherical limit of matrix models \cite{Kri92,Dub91}
as well as of
LGM it is known that the dynamics over several first Miwa times and $p$-times
coincide. As this dynamics, besides integrable equations, is consistent with
the
string equation which is a first-derivative equation, one can expect that the
derivatives of the (proper re-defined) $\tau $-function with respect to both
Miwa and $p$-times are equal. This is really the case {\it in all genera}, and
the main statement we are going to prove reads
\beq\label{4}
Z_{GKM}[V|T_k] = \tau _V[T_k] =
\exp \left( - {1\over 2}\sum    A_{ij}(t)(\tilde T_i+t_i)(\tilde T_j+t_j)
\right)  \tau _p[\tilde T_k+ t_k],
\eeq
where
\footnote{Proper
changing the variable $X$ in the matrix integral, one can always
choose $v_{p+1}=1$, $v_p=0$, what is convenient normalization and usually
implies throughout the paper.
}
\beq\label{F1}
V(x) = \sum ^{p+1}_{k=1} {v_k\over k} x^k,
\eeq
\beq\label{5}
\tilde T_k = {1\over k}Tr \tilde M^{-k},
\eeq
\beq\label{F2}
\tilde M^p = V'(M) \equiv  W(M),
\eeq
\beq\label{F3}
A_{ij} = Res  W^{i/p} dW^{j/p}_+ =
{\partial ^2\log \ \tau_{0} \over \partial t_i\partial t_j},
\eeq
$\tau _0$ is the quasiclassical $\tau $-function and parameters  $\{t_k\}$  are
certain linear combinations of the coefficients  $\{v_k\}$  of the potential
\beq\label{6p}
t_k = {p\over k(p-k)}Res  W^{1-k/p}(\mu ) d\mu
\eeq
$i.e.$ that a generic GKM $\tau $-function is expressed through $\tau
$-function of
$p$-reduction, depending only on the sum of time-variables $\tilde T_k$ and
$t_k$, associated with deformations in ``$q$-" and ``$p$-directions"
respectively. The change of the spectral parameter in (5)  $M \rightarrow
\tilde M$  (and corresponding times  $T_k \rightarrow  \tilde T_k$)  is a
natural step from the point of view of equivalent hierarchies. As to original
GKM, its $p$-dependence is not seen in eq.(1), though it can be also introduced
implicitly through the choice of integration contour.

We omitted from this letter some proofs which will be published in the separate
publication \cite{KMMM92c},
together with more detailed discussion of some points
discussed in the last section of this letter.

\bigskip

{\bf 2.} {\it Time derivatives}.  The first step of our calculations concerns
the derivatives of  $Z_{GKM}$ with respect to the time-variables  $T_k$. Such
derivatives define nonperturbative correlators in string models and are of
their own interest for the theory of GKM. The derivatives with respect to
$T_k$ with  $k \geq  p+1$  (responsible for the correlators of irrelevant
operators) are not very easy to evaluate, things are simpler for  $T_k$ with
$1 \leq  k \leq  p$. This is due to the fact mentioned in the previous section
that these derivatives should be simple expressed through derivatives with
respect to $p$-times. Using the obvious notation of average so that  $Z_{GKM} =
\left< 1\right> $, we have
\beq\label{6}
\left.{\partial Z_{GKM}\over \partial T_k}\right| _V = \left< TrM^k -
TrX^k\right> , \ \ \  1 \leq  k \leq  p
\eeq
It is implied that the derivative in the $l.h.s.$ is taken under constant
values of  $v_m$, $i.e.$ preserving the form of the potential  $V$.

The $r.h.s$. of (\ref{6}) can be also represented as
\beq\label{7}
- {\partial Z_{GKM}\over \partial T_k}\left| _V =
\left< Tr{\partial V(X)\over \partial v_k} -
Tr{\partial V(M)\over \partial v_k}\right> \hbox{, }   1 \leq  k \leq
p\right.
\eeq
which looks similar but actually {\it is different} from  $-
{\partial \over \partial v_k}Z_{GKM}$, as it would be if (\ref{4}) does
not contain $t$-dependence of the coefficients $A_{ij}$ of the quadratic form.
The problem is that  ${\partial \over \partial v_k}Z_{GKM}$ gets contributions
not only from differentiating $V(X) - V(M)$ in exponentials in
(\ref{1}) but
also from the term  $V'(M)(X-M)$ as well as from the pre-exponential
$C[V|M]$. Let us keep this in mind for a while and now we turn to a slightly
different question.

\bigskip

{\it Change of $M$-variables}.  Another question which can be asked about
$Z_{GKM}[V|T]$  is whether it corresponds to anything reasonable if the
KP-times are introduced in a way different from (\ref{3}). For example, let us
{\it define}, instead of (\ref{5}), new variables
\beq\label{8}
T^{(f)}_k = {1\over k}Tr[f(M)]^{-k}
\eeq
If  $f(m) = m(1 + o(1/m))$  this can be considered as an allowed change of
spectral parameter and it is not difficult to preserve the statement that
$Z_{GKM}$ is a KP $\tau $-function of  $T^{(f)}_k$ -- variables: in fact, it is
enough to modify slightly the definition of  $C[V|M]$. Instead of
(\ref{2}) it should be now
\beq\label{9}
C^{(f)}[V|M] = C[V|f(M)].
\eeq
Then
\beq\label{10}
Z^{(f)}_{GKM} \equiv  {C^{(f)}\over C}Z_{GKM} = \tau (T^{(f)}_k).
\eeq
Of course, this procedure changes also the point of the Grassmannian.

The next step will be to find out an analog of the results
(\ref{6}) and (\ref{7}) with
$T_k$'s  replaced by  $T^{(f)}_k$'s. It is easy to do using eq.(\ref{6})
\bea\label{11}
\left.{\partial \over \partial T^{(f)}_k}Z_{GKM}\right|
_V = \sum  _l
{\partial T_l\over \partial T^{(f)}_k}{\partial Z_{GKM}\over \partial T_l} =
\left< \sum  _l {\partial T_l\over \partial T^{(f)}_k}(TrX^l -
TrM^l)\right> =\nn\\
= \left< Tr[f^k_+(X)] - Tr[f^k_+(M)]\right> \hbox{, }    1
\leq  k \leq  p,
\eea
where we used the following transformation of times:
\beq\label{n1}
T_k^{(f)} = {1\over k}\sum  _j jT_j
Res \ \lambda ^{j-1}f^{-k}(\lambda )d\lambda ,
\eeq
\beq\label{n2}
T_k = \sum  _j T_j^{(f)}
Res \ \lambda ^{-k-1}f^j(\lambda )d\lambda .
\eeq

Note that in the $l.h.s.$ of (\ref{11}) we differentiate  $Z_{GKM}$ and not
$Z^{(f)}_{GKM}$, in the latter case we get an additional correction of
$\partial /\partial T[log C^{(f)}/C]$  which is a sort of ``quantum"
correction,
since  $C[V|M]$  can be considered as a quantum correction to the classical one
given by terms in the exponentials
\footnote{The
obvious notation is  $\left< \ldots \right> _f \equiv
{C^{(f)}\over C}\left< \ldots \right> $,  $e.g.$  $Z^{(f)}_{GKM} =
\left< 1\right> _f$.}.

Now, in order to find the analog of (\ref{7}) let us introduce new parameters
$v^{(f)}_k$ in such a way that
\beq\label{12}
{\partial V(X)\over \partial v^{(f)}_k} = f^k_+(X)
\eeq
With such definitions (\ref{11})
takes the form
\bea\label{13}
\left.- {\partial \over \partial T^{(f)}_k}Z_{GKM}\right|
_V = \left< Tr{\partial V(X)\over \partial v^{(f)}_k} -
Tr{\partial V(M)\over \partial v^{(f)}_k}\right>,\nn\\
\left.\left.- {\partial \over \partial T^{(f)}_k}Z^{(f)}_{GKM}\right|
_V = \left< Tr{\partial V(X)\over \partial v^{(f)}_k} -
Tr{\partial V(M)\over \partial v^{(f)}_k} \right> -
{\partial \over \partial T^{(f)}_k}\log {C^{(f)}\over C}\right| _V.
\eea
Again, in general the $r.h.s.$ of (\ref{11}) is not the same as
\bea\label{14}
- {\partial \over \partial v^{(f)}_k}Z^{(f)}_{GKM} =
\left< Tr{\partial V(X)\over \partial v^{(f)}_k} -
Tr{\partial V(M)\over \partial v^{(f)}_k} \right>_f +
Tr\left[{\partial V'(M)\over \partial v^{(f)}_k}\left< X - M\right> _f\right]
-\nn\\
- {\partial \log C\over \partial v^{(f)}_k}\left< 1\right> _f
\eea

Now, let us allow the matrix  $M$  itself to change when  $v^{(f)}_k$ are
varied, this gives an additional contribution to (\ref{14}) of the form
\beq\label{15}
TrV''(M){\partial M\over \partial v^{(f)}_k}{\left< X - M\right>}_f  -
Tr{\partial M\over \partial v^{(f)}_k}
{\partial logC\over \partial M}\left< 1\right> _f
\eeq
The second term in the $r.h.s.$ of (\ref{14}) and the first term in
(\ref{15}) can be now combined into
\beq\label{16}
{Tr \left[{\partial V'(M)\over \partial v^{(f)}_k} +
V''(M){\partial M\over \partial v^{(f)}_k}\right]{\left< X - M\right>}_f  = Tr
{dV'(M)\over dv^{(f)}_k}{\left< X - M\right>}_f }
\eeq
and we conclude that
\beq\label{17}
\left.\left( {\partial \over \partial T^{(f)}_k} -
{\partial \over \partial v^{(f)}_k}\right) Z^{(f)}_{GKM}\right|_{T_k^{(f)}=0}
= 0,
\eeq
only if the expression (\ref{16}) is equal to zero and all Miwa times
$\tilde T_k^{(f)}=0$. The latter requirement is the direct consequence of the
fact
that
all normalization contributions are bilinear or linear forms of Miwa times
(this point is discussed in details in the section 4).

\bigskip

{\bf 3.} $p${\it -times.} Now let us consider what means that there are no
corrections to (\ref{17}) or the contribution of (\ref{16})
is equal to zero. It implies first, that  $V'(M)\equiv W(M)$  is a fixed
function of the new variable $\tilde M \equiv f(M)$, and second, the
leading degree of this function is
$p$ (to dive asymptotic expansion of  $f(M)$). Thus, it allows one to
choose  $f(M)$ in the monomial form of degree $p$:
\beq\label{18}
W(M) = f(M)^p \equiv  \tilde M^p.
\eeq

Now, provided (\ref{16}) is equal to zero, one obtains:
\beq\label{19}
{\partial \over \partial T^{(f)}_k} \log Z_{GKM} =
{\partial \over \partial v^{(f)}_k}\log Z^{(f)}_{GKM}.
\eeq

This is almost the relation we need. Below, we will explain how the partition
function can be redefined in order to have the equality of derivatives like
(\ref{19})
having the same objects in both sides of the equation.

Thus, we are led to special time variables induced by a special transformation
of the spectral parameter $\mu \rightarrow f(\mu ) = W(\mu )^{1/p}$. These
$p$-times
are just those appeared in the paper \cite{DVV91b} in the context of
Landau-Ginzburg topological theories, and in the papers \cite{Kri92,Dub91}
in the
framework of quasiclassical (or dispersionless) hierarchies. We are going to
demonstrate that these times
are also natural in our approach and acquire a nice interpretation. Indeed,
the explicit expression,
\beq\label{20}
t_k = {p\over k(p-k)} Res W^{1-k/p} d\mu
\eeq
can be easily continued to negative values of the index $k$ (the
negative times will be discussed in detailed publication \cite{KMMM92c},
see also the sect.6).
Then we get two following formulas:
\beq\label{21}
\mu  = - {1\over p}\sum ^{p+1}_{-\infty } kt_k \tilde \mu ^{k-p},
\eeq
\beq\label{22}
V(\mu ) - \mu V'(\mu ) = \sum ^{p+1}_{-\infty } t_k \tilde \mu ^k.
\eeq
The first of these formulas can be easily modified for any variables
$v^{(f)}_k$,
but the second one is specific for $p$-times and implies the natural
interpretation of the exponential pre-factor in eq.(\ref{1})
as the standard essential
singularity factor in the Baker-Akhiezer function of $p$-time variables.

This can be done more transparently by the following procedure. Let us
consider
the equation for the Baker-Akhiezer function with all times  $T_k=0$ except for
$T_1=x$ \cite{KMMMZ91b}:
\beq\label{23}
\ [W(\partial )+x]\Psi (\mu ,x) = W(\mu )\Psi (\mu ,x).
\eeq
One can look at this equation from two different points of view. On one hand,
it
can be considered as an initial boundary condition for the standard KP
($V'$-reduced)
dynamics over Miwa times. Then this dynamics is described by the isospectral
deformations and can be transformed to the standard KP dynamics by the
re-expressing the spectral parameter $\mu $ through a pseudo-differential
operator using (\ref{23}):
\beq\label{24}
\mu \Psi (\mu ,x) = [\partial  + \sum _{i=1}^{\infty}
u_{i+1}\partial ^{-i}]\Psi (\mu ,x) \equiv L \Psi(\mu ,x).
\eeq
Then (\ref{23}) is a statement about reduction -- $i.e.$,
{\it pseudo}-differential operator (namely $W(L)$ ) is the {\it differential}
one.

On the other hand, the eq.(\ref{23}) can be considered as describing the
(consistent) dynamics over $p$-times, with zero Miwa times and $p$-th KdV
initial boundary condition:
\beq\label{25}
\ [\partial ^p+x]\Psi (\mu ,x) = \mu ^p\Psi (\mu ,x).
\eeq
This dynamics corresponds to a special {\it non}-isospectral deformations in
$p$-times and describes the
flows between different reductions (or equivalent hierarchies, see the sect.4)
and {\it a priori} has nothing to do with the KP hierarchy-structure of GKM.

Thus, it turns out that these two dynamics are essentially the same in a
special (matrix model) point of the Grassmannian determined by string equation.
This amusing fact was previously proved in spherical limit in the papers
\cite{Kri92,Dub91},
and will be explained in details in the sect.4-5. In other words the exact
(non-perturbative)
solution in this
special point of the Grassmannian equals to a quasiclassical one, $i.e.$
the quasiclassical approximation
is in a sense exact. Leaving the correct formulation of this statement to
sect.4--5,
now we are going to discussion of the quasiclassical hierarchies and to
demonstrate,
what is remarkable in the chosen point of the Grassmannian.

\bigskip

{\it Quasiclassical hierarchies.}  Let us see what is specific in eq.(\ref{23})
from general point of view of integrable theories. The
point of the Grassmannian is determined by the set of coefficient functions of
Lax
operator of $p$-reduced KP hierarchy. In our case, this is given by the
$l.h.s.$ of the eq.(\ref{23})
and its defining property is that it {\it does not} depend on
$x$ except for the constant term ($i.e.$ $Res_\partial
\tilde W\partial ^{-1})$ depends linearly on $x$. Now let us look at the KP
hierarchy with Lax operator obeying this property.

The Lax representation of $p$-reduced KP hierarchy is \footnote{It is
generally rather natural to consider pseudo-differential operators in the
context of GKM, since $\left< W_+^{k/p}(X)\right>$ can be simply rewritten
as $\ll W_+^{k/p} (\partial)\gg$, where $\ll \ldots \gg$ implies that the
operator stands under integral in (\ref{1})
in front of the term $exp{Tr\Lambda X}$,
$\Lambda \equiv V'(M)$, $\partial \equiv \partial /\partial \Lambda$.}
\beq\label{26}
{\partial \tilde W(\partial )\over \partial t_k} =
[\tilde W^{k/p}_+(\partial ),\tilde W(\partial )],
\eeq
where $\tilde W(\partial )$ is a differential operator
$\partial \equiv \partial /\partial x$, and the coefficients of the operator
are
allowed to depend on the first time $x$ as well as on other times $t_k$.
Particular solution (or specific point of the Grassmannian) defining matrix
models is distinguished by the requirement that all except for the constant
term in $\tilde W$ are, in fact, $x$-independent. This is a consistent
requirement provided only by the linear $x$-dependence of this constant term,
$i.e.$
\beq\label{27}
\tilde W(\partial ) = W(\partial ) + x,
\eeq
where $W(\partial )$ at the $r.h.s.$ is supposed to have no $x$-dependence at
all. Then  $(W(\partial )+x)^{k/p}_+$ does not depend on $x$ for
$1\leq k\leq p-1$, and just equals to $W^{k/p}_+(\partial )$, so that the
commutator at the $r.h.s.$ of (\ref{26})
acquires the only contribution from $x$-term
in $\tilde W$, and (\ref{26})
turns exactly into the equation \footnote{Of course, it is also possible
to obtain this equation immediately by taking a
$\mu$-derivative of the eq.(\ref{12}).}
\beq\label{28}
{\partial W(\lambda )\over \partial t_k} =
{\partial W^{k/p}_+(\lambda )\over \partial \lambda }\hbox{, }
1\leq k\leq p-1,
\eeq
where $\lambda $ is a formal parameter.

This hierarchy (\ref{28}), indeed, was obtained as a quasiclassical (or
dispersionless) hierarchy satisfying the string equation
\cite{Kri92,Dub91,TT92}. However, the same equations (\ref{28})
are exactly equivalent to the KP hierarchy on particular class of solutions
(of the type of (\ref{27}). Thus, we state that the
specific matrix model point of the Grassmannian gives rise to the dynamics with
respect to the first $p$-times (if all Miwa times are equal to zero) which is
simultaneously KP and dispersionless dynamics, $i.e.$ the quasiclassical
approximation is exact. Moreover, this dynamics is the same as the dynamics
with respect to the first Miwa times. Therefore, we reproduce the result
(\ref{17}) in other terms.

\bigskip

{\bf 4.} {\it Equivalent hierarchies.}
In this section we would like to discuss in
details the general framework of integrability for the problems discussed
above,
in particular the notion of equivalent hierarchies.

The notion of equivalent solution of the KP hierarchy was introduced in
\cite{Shi86}
and was based on the particular transformations of the time variables.
This concept was further developed in \cite{Tak91} for the
general Zakharov-Shabat
equations and the Toda lattice hierarchy (the KP case, as cited in \cite{Tak91}
, was considered in the unpublished paper by M. Noumi). Let us consider the
general Zakharov-Shabat system
\beq\label{4.1}
\partial B_i/\partial T_j - \partial B_j/\partial T_i + [B_i,B_j] = 0,
\eeq
where Hamiltonians $B_i$ are the differential polynomials of $i$-th degree
and are {\it not restricted} generally to be  $(L^i)_+$, where $L$ is a
pseudo-differential operator (\ref{24}) giving a solution to the KP hierarchy.
Then the system (\ref{4.1})
contains the equations of the KP hierarchy as the subset since
in general the first two polynomials have the form $B_2 = \partial ^2 + 2u_2$,
$B_3 = \partial ^3 + 3(u_2 + a) + 3b$ with {\it three} independent functions
while in the KP case $B_2 = \partial ^2 + 2u_2$ , $B_3 = \partial ^3 + 3u_2 +
3(u_3 + u_{2,x})$ have only {\it two} independent functions. At the same
time the system (\ref{4.1})
contains more equations which restrict the functional
dependence of the additional functions on the  time variables. For example,
$a(T)$ is $x$-independent and, therefore, $u_2$ satisfy the usual
Kadomtsev-Petviashvili equation. This freedom is the reflection of the fact
that zero curvature form is covariant under the arbitrary upper-triangle
(gauge) transformations of the times
\beq\label{S.8}
T_i \rightarrow  \tilde T_i = T_i + \xi _i(T_{i+1}\hbox{, } T_{i+2}\hbox{,
...)}
\eeq
and therefore the new differential operators defined by
\beq\label{S.9}
B_i(T) = \sum  _j {\partial \tilde T_j\over \partial T_i} \tilde B_j(\tilde T)
\eeq
also satisfy (\ref{4.1}).
In this sense the given time transformation (\ref{S.9}) defines
the equivalent hierarchy. The covariance of (\ref{4.1})
under the transformation (\ref{S.8})
gives a possibility to eliminate the functional freedom in the definition of
the polynomials $B_i$. Indeed, in \cite{Tak91} it has been proven that for an
arbitrary
$B_i$ there exists the unique transformation (\ref{S.8})
such that $\tilde B_i$
determine the KP hierarchy, $i.e.$ $\tilde B_i$ can be represented in the form
of $(L^i)_+$ for some operator $L$ and satisfy the Lax equations:
\beq\label{4.2}
\partial L/\partial T_i = [B_i,L].
\eeq
Only in this case the solution of these equations can be described by a single
$\tau $-function.

Here we consider only the very restricted class of above transformations which
are induced by the variation of the spectral parameter $\mu $ and,
thus, still preserve the notion of the $\tau $-function. Let us introduce an
arbitrary function $f(\lambda )$ which is expandable in the formal Laurent
series $f(\lambda ) = \Sigma f_i\lambda ^i$ ($f_1\equiv 1$) and perform the
transformation of the spectral parameter
\beq\label{S.10}
\tilde \mu  = f(\mu )
\eeq
or, equivalently, define a new Lax operator
\beq\label{S.11}
\tilde L = f(L) \equiv  L +\sum ^0_{i=-\infty } f_iL^i
\eeq
We should note that the transformation (\ref{S.10})
respects the KP structure, $i.e.$ maps
the given KP hierarchy onto equivalent one:
\beq\label{S.12}
{\partial \tilde L\over \partial \tilde T_i} = [\tilde B_i\hbox{, }
\tilde L]\hbox{  , }   \tilde B_i= (\tilde L^i)_+
\eeq
where new times $\tilde T$ are introduce by the eq.(\ref{8}) and
\beq\label{S.14}
\tilde B_i(\tilde T) = \sum  _j {\partial T_j\over \partial \tilde T_i} B_j(T).
\eeq
Now let us consider the $\tau $-function given in Miwa variables. It can be
represented in the determinant form \cite{KMMMZ91b}
\beq\label{S.7}
\tau (T) = {\det \ \phi _i(\mu _j)\over \Delta (\mu )},
\eeq
where times $\{T\}$ are parameterized in the Miwa form (\ref{3}) and
$\{\phi _i(\mu )\}$ are the basic vectors determining the point of Grassmannian
\cite{SW85}.

The relation between $\tau $-functions of the equivalent hierarchies can be
easily derived from the eq.(\ref{S.7}) by an identical transformation:
\beq\label{S.15}
\tau (T) = {\Delta (\tilde \mu )\over \Delta (\mu )}
\prod_i
[f'(\mu _i)]^{1/2} \tilde \tau (\tilde T)
\eeq
where $\tilde \tau (\tilde T)$ as function of times $\tilde T$
has the determinant
form (\ref{S.7}) with the basic vectors
\beq\label{S.16}
\tilde \phi (\tilde \mu ) =
[f'(\mu (\tilde \mu ))]^{1/2}\phi _i(\mu (\tilde \mu ))
\eeq
By a direct calculation one can show that pre-factor in
eq.(\ref{S.15}) may be
represented in the form
\beq\label{S.17}
{\Delta (\tilde \mu )\over \Delta (\mu )}
\prod_i
[f'(\mu _i)]^{1/2} = \exp \left( - {1\over 2}
\sum _{i,j}A_{ij}\tilde T_i\tilde T_j\right)
\eeq
where
\beq\label{S.18}
A_{ij} = Res\ f^i(\lambda )d_\lambda f^j_+(\lambda ).
\eeq
The notion of equivalent hierarchies is very useful in the context of GKM. In
\cite{KMMMZ91b}
we proved that the $\tau $-function of GKM corresponds to $V'$-reduced KP
hierarchy (in the case of polynomial $V(\lambda )$), $i.e.$ $V'(L)$ is a
differential operator. Therefore, it is reasonable to consider the
transformation (\ref{S.10}) with
\beq\label{S.19}
f(\mu ) = [V'(\mu )]^{1/p} \equiv [W(\mu )]^{1/p}.
\eeq
In this case the equivalent hierarchy determined by the operator
$\tilde L = V'(L)$ is $p$-reduced one and we can treat the partition function
of GKM in the terms of $p$-reduced $\tau $-function $\tilde \tau (\tilde T)$
with the suitable deformations. These deformations are of two kinds. The first
deformation corresponds to the transformation of times $T_i \rightarrow
\tilde T_i$ (see (\ref{n1}),(\ref{n2}));
the second one corresponds to the multiplication of $p$-reduced
$\tau $-function by the exponential pre-factor which is quadratic in times
$\{\tilde T\}$ and depends non-linearly on the coefficients of the potential
$V(\lambda )$. In the series of papers \cite{Kri92,TT92} it was shown that the
matrix $A_{ij}$ determined by the eqs.(\ref{S.18}), (\ref{S.19})
can be represented in the form
\beq\label{S.22}
A_{ij} = \left( {\partial ^2\over \partial t_i\partial t_j}
\log \ \tau _0(t)\right) _{t_{p+2}=...=0} i\hbox{, } j = 1,2, ...
\eeq
where $\tau _0(t)$ is the $\tau $-function of the quasiclassical $p$-reduced KP
hierarchy restricted on ``small phase space" \cite{DW90}.
Here $\{t\} = \{t_1$, ... ,
$t_{p-1}$, $t_{p+1} = - {p\over p+1}\}$ are $p$-times determining by
eq.(\ref{20})
and corresponding quasiclassical ``Lax operator" has the form
\beq\label{S.23}
{\cal L}(\lambda ) = [W(\lambda )]^{1/p}\hbox{ ;}
\eeq
this is essentially corresponds to the transformation of the spectral parameter
(\ref{S.19}).
Thus we can see that $\tau $-functions of the equivalent hierarchies
(which are induced by variation of the spectral parameter) can be transformed
to each other along quasiclassical flows.

We should remark also that the quasiclassical $\tau $-function satisfies the
homogeneity condition \cite{Kri92,TT92}
\beq\label{S.24}
\sum  _i t_i {\partial \over \partial t_i} \log \ \tau _0(t) =
2\ \log \ \tau _0(t).
\eeq

\bigskip

{\bf 5.} {\it Main results.} From the eqs.(\ref{S.15})-(\ref{S.19})
we see that
\beq\label{S.25}
\tau (T(\tilde T)) = \tilde \tau (\tilde T) \exp \left( - {1\over 2}
\sum _{i,j}A_{ij}\tilde T_i\tilde T_j\right)
\eeq
where $A_{ij}$ is determined through the quasiclassical $\tau $-function
$\tau _0(t)$ (eq.(\ref{S.22})). Let us introduce by definition the new
$\tau $-function $\hat \tau (\tilde T)$ of the $p$-reduced KP hierarchy:
\beq\label{S.26}
\tilde \tau (\tilde T) \equiv  {\hat \tau (\tilde T)\over \tau _0(t)}
\exp \left( \sum  _j jt_{-j}\tilde T_j\right) ,
\eeq
where
\beq\label{S.27}
{{\hat \tau (\tilde T)} \over {\tau _0(t)}} = {\det
\hat \phi _i(\tilde \mu _j)\over \Delta (\tilde \mu )}
\eeq
and the point of the Grassmannian is determined now by the basic vectors
\beq\label{S.28}
\hat \phi _i(\tilde \mu ) = [p\tilde \mu ^{p-1}]^{1/2} \exp \left(
\sum ^{p+1}_{j=1}t_j\tilde \mu _j\right)  \int
x^{i-1}e^{V(x)-x\tilde \mu ^p}dx.
\eeq
Then it is easy to show that $\hat \tau (\tilde T)$ satisfies the $L_{-1}$-
constraint with shifted KP-times$:$
\beq\label{S.29}
\sum ^{p-1}_{k=1}k(p-k)(\tilde T_k+t_k)(\tilde T_{p-k}+t_{p-k}) +
\sum ^\infty _{k=1}(p+k)(\tilde T_{p+k}+t_{p+k}){\partial \over \partial %
\tilde T_k} \log  \hat \tau (\tilde T) = 0 ,
\eeq
where $t_i$ are equal to zero for $i \geq  p+2$ (see eq.(\ref{20})) so
$\hat \tau (\tilde T)$ actually depends only on the sum of (``quantum") Miwa
times and (quasiclassical) $p$-times. Moreover, from eqs.(\ref{19})
and (\ref{S.25}), (\ref{S.26}) it directly follows that
\beq\label{S.30}
\left( {\partial \over \partial \tilde T_i} -
{\partial \over \partial t_i}\right) \hat \tau (\tilde T) = 0\ \ \ \ \ i =
1,2, ... ,p-1.
\eeq
Therefore, our final answer is that in the polynomial case $(V(X) =
\sum ^{p+1}_{k=1} {v_k\over k} X^k)$ one can represent the GKM partition
function in the following form:
\beq\label{S.31}
\tau (T) = \hat \tau (\tilde T + t)\exp \left( - {1\over 2}
\sum _{i,j}\left( {\partial ^2\over \partial t_i\partial t_j}
\log \ \tau _0(t)\right) (\tilde T_i+t_i)(\tilde T_j+t_j)\right)
\eeq
(in order to obtain the last formula we have used the homogeneity condition
(\ref{S.24})),
where $\hat \tau (\tilde T+t)$ is the standard $\tau $-function
corresponding to monomial $V(x) = {X^{p+1}\over p+1}$ and
\beq\label{S.32}
\hat \tau (t) = \tau _0(t)
\eeq
is the $\tau $-function of the quasiclassical $p$-reduced KP hierarchy with the
``Lax operator" (\ref{S.23})
(in fact, in the second part of the section 3 we have
already met this property that the quasiclassical function is exact when Miwa
times are equal to zero).

As $\hat \tau $ satisfies the standard string equation for multi-matrix model,
one can also write it as GKM integral with the monomial potential
\cite{KMMMZ91a,KMMMZ91b}, but
with $\tilde T+t$ playing the role of Miwa times. Therefore, $\hat \tau $
describes simultaneously $(p-1)$-matrix model, and there is no smooth
transition from this to another $p$. Thus, $\hat \tau $ is singular under
changing $p$, as well as exponential in (\ref{S.31}).
At the same time the complete
answer corresponding to GKM integral is regular, and this is one of main
advantages of GKM approach.

\bigskip

{\bf 6.}{\it Discussion}.
In this letter we have proposed a way how two {\it a priori}
different objects like  $N=2$  topological LG theories and conformal  $c<1$
minimal models coupled to Polyakov $2d$ gravity can be unified in the framework
of GKM, exploiting integrable structure appeared in these objects. However,
this connection deserves further understanding.

The main formula (\ref{4})
describes the interpolating flow between two $(p,q)$ and
$(p',q)$ models in terms related to LGM. Of course, this is only a {\it local}
description around a given ``critical point". It shows that the flow has a
complicated ``phase"-structure with the LG theory being responsible for its
first phase --- on a particular  $p$-orbit determined by the order of the
potential  $V'(X) \equiv  W(X) = X^p+...$

Formula (\ref{4}) demonstrates that such flow can be more or less absorbed
into the redefinition of times:  $T \rightarrow  \tilde T$  by
\beq\label{*}
\tau _p(T) \rightarrow  {\tau _p(\tilde T + t)\over Z_{cl}(\tilde T+t|t)}.
\eeq
Here $Z_{cl}$ denotes the classical contributions to the partition function,
the $\tau $-function in the numerator corresponds to the same point of the
Grassmannian (satisfies the same string equation) and the parameters of the
flow just add to the corresponding KP times. It means that the LG flow is
generated by first  $p$  primaries of minimal conformal model plus $2d$ gravity
theory being equivalent to the primary LG fields.

On the ``boundary" of this phase $p$-times diverge thus being non-adequate
parameters for the description of the ``phase"-transition between different
$p$-orbits, what is typical for the phase transitions.

The ``classical" partition function in the denominator of (\ref{*})
has to be better understood from the LG point of view. The derivative
\beq\label{**}
A_{ij}(t) =  {\partial ^2\over \partial t_i\partial t_j}\log \ \tau _0 =
Res \ W^{i/p} dW^{j/p}_+
\eeq
(see also \cite{TT92}) is an object of similar nature to those appearing
in LGM on
the so called Grothendick residue formulas. Indeed, the scalar product
defining the Grothendick residue \cite{Vaf90,DVV91b}:
\beq\label{***}
\left< \phi _i\phi _j\right>  = Res
{\phi _i\phi _j\over W'} d\mu
\eeq
together with
\beq\label{****}
\phi _i(\mu ) \equiv  {\partial \over \partial t_i}W(\mu ) =
{\partial \over \partial \mu } W^{i/p}_+
\eeq
gives
\beq\label{*5}
\left< \phi _i\phi _j\right>  \sim  Res\ W^{(i-p)/p} dW^{j/p}_+
\eeq
which is formally
\beq\label{*6}
{\partial ^2\over \partial t_{i-p}\partial t_j}\log \ \tau _0
\eeq
if we introduce negative $p$-times. The negative times should correspond
to a deformation under arbitrary (non-polynomial)
change of spectral parameter  $\tilde \mu  =
f(\mu )$  though the integrable structure (with respect to $p$-times $\{t\})$
is not yet completely clear.

The basic feature of topological theories is ring structure \cite{LVW89}. In
our case a sort of ring appears immediately from the reduction condition
\beq\label{A}
W(\mu )\varphi _i(\mu ) = \sum  _j C_{ij}\varphi _j(\mu )
\eeq
with $\mu $-independent  $C_{ij}$ which is a direct consequence of generalized
Airy equation \cite{KMMMZ91b}.  In new basis the last relation turns to be
\beq\label{B}
\tilde \mu ^p\tilde \varphi _i(\tilde \mu ) = \sum  _j
\tilde C_{ij}\tilde \varphi _j(\tilde \mu )
\eeq
and the only difference with conventional LGM is that operations (\ref{A}) and
(\ref{B})
are determined on non-polynomial functions. It also deserves noting that the
Grothendick residue formula (\ref{***})
acquires an especially simple form in the basis
\beq\label{C}
\Phi _i(\mu ) = W'(\mu )\phi _i(W(\mu )).
\eeq
Indeed
\bea\label{D}
\left< \Phi _i\Phi _j\right>  = Res
{\Phi _i(\mu )\Phi _j(\mu )\over W'(\mu )} d\mu= Res
W'(\mu )\phi _i(W(\mu ))\phi _j(W(\mu )) d\mu =\nn\\
= Res \phi _i(W)\phi _j(W)dW
\eea
which is also natural in the framework of topological theories \cite{Los92}.

Now let us stress that we can not immediately identify a concrete object
from the GKM theory with the
LGM partition function. This is due to the fact that there is still a little
information about the latter one. Moreover, only its $p$-time dependence is
known while the dependence on Miwa times is much more subtle point.
It can be determined only having explicit expressions for{\it all} correlators,
but there are no formulas for 5-point and higher correlators even of the
primary fields (see \cite{Los92}). Therefore, the LGM partition function at the
moment can
be identified with GKM objects only hypothetically. It is reasonable to
conjecture that this partition function might coincide with $\hat Z$. This is
consistent with the fact that LGM can not smoothly interpolate between
different $p$ orbits.

Thus, we see that there exists a deep connection between the structure of LGM
and matrix model formulation of $2d$ gravity which should be revealed better
and we are going to return to this
problem in the separate publication \cite{KMMM92c}.

\bigskip

We are grateful to B.Dubrovin, A.Lossev, I.Krichever, M.Olshanetsky,
T.Takebe, S.Theisen, C.Vafa and A.Zabrodin for illuminating
discussions. A.Mar. is grateful for warm hospitality to Lyman Laboratory of
Harvard University, Physics Department of University of British Columbia and
Physics Department of California Institute of Technology where preliminary part
of the work has been done.

\end{document}